\documentclass[useAMS]{mn2e}

\usepackage{amssymb,amsmath,psfig,times}
\def\gsim{ \lower .75ex \hbox{$\sim$} \llap{\raise .27ex \hbox{$>$}} }
\def\lsim{ \lower .75ex\hbox{$\sim$} \llap{\raise .27ex \hbox{$<$}} }

\def\sc{Schwarzschild}

\title[PMN J2345--1555: from red to blue]
{The red blazar PMN J2345--1555 becomes blue}  
\author[G. Ghisellini,  et al.] 
{G. Ghisellini\thanks{Email: gabriele.ghisellini@brera.inaf.it}, 
F. Tavecchio, L. Foschini, G. Bonnoli, G. Tagliaferri \\
$^1$ INAF -- Osservatorio Astronomico di Brera, Via Bianchi 46, I--23807 Merate, Italy\\
}
\begin{document}  

\maketitle

\begin{abstract}
The Flat Spectrum Radio Quasar PMN J2345--1555 is a bright $\gamma$--ray source,
that recently underwent a flaring episode in the IR, UV and $\gamma$--ray bands.
The flux changed quasi simultaneously at different frequencies, suggesting that
it was produced by a single population of emitting particles, hence
by a single and well localized region of the jet.
While the overall Spectral Energy Distribution (SED) before the flare was
typical of powerful blazars (namely two broad humps peaking in the far IR and below 100 MeV
bands, respectively), during the flare the peaks moved to the optical--UV and 
to energies larger than 1 GeV, to resemble low power BL Lac objects, 
even if the observed bolometric luminosity
increased by more than one order of magnitude.
% The two humps moved to the optical--UV and to energies larger than 1 GeV.
We interpret this behavior as due to a change of the location of the emission region
in the jet, from within the broad line region, to just outside.
The corresponding decrease of the radiation energy density as seen in the comoving
frame of the jet allowed the relativistic electrons to be accelerated to higher energies, and
thus produce a ``bluer" SED.
\end{abstract}

\section{Introduction}

PMN J2345--1555 is a Flat Radio Spectrum Quasar (FSRQ) at $z=0.621$ (Healey et al. 2008),
bright in $\gamma$--rays, and detected by the Large Area Telescope (LAT)
onboard the {\it Fermi} satellite in the first 3 months of its operation
(Abdo et al. 2009), and later present in the 1LAC (catalog of AGN detected
in the first 11 months of operations, Abdo et al. 2010) and 2LAC (first 2 years,
Nolan et al. 2012).
The $\gamma$--ray spectra were described by simple power laws.
The fluxes above 100 MeV are reported
in Tab. \ref{summary}, together with the photon spectral indices. 
It can be seen that the source varied on long timescales, even if the reported
fluxes are averaged over the entire period of observations.
On the other hand the spectral index $\Gamma$ was always greater than 2, 
indicating a $\nu F_\nu$ peak at energies smaller than 100 MeV.
This is typical for the FSRQ class (see Abdo et al. 2010; Ghisellini, Maraschi \& Tavecchio 2009).

% with a flux above 100 MeV of $(10.3\pm1.3)\times 10^{-8}$. 
% Assuming a power law spectrum, the photon index was $\Gamma=2.42\pm 0.12$.
% It was present in the 1LAC AGN sample of {\it Fermi} (Abdo et al. 2010) listing
% the AGNs detected after 11 months of operations,
% with a [1--100 GeV] flux of $(1.5\pm0.3)\times 10^{-9}$ ph cm$^{-2}$ s$^{-1}$ and
% a slope $\Gamma=2.37\pm 0.1$. 
% With these parameters the flux above 100 MeV would be $3.5\times 10^{-8}$ ph cm$^{-2}$ s$^{-1}$,
% about a factor 3 below the flux during the first 3 months, but with the same slope
% (within the errors).
% In the 2LAC AGN catalog (Nolan et al. 2012), comprising all {\it Fermi} detected AGN after 2 years of operations,
% the source was listed with an average [1--100 GeV] flux of $(4.2\pm 03)\times 10^{-9}$ ph cm$^{-2}$ s$^{-1}$
% and $\Gamma=2.19\pm 0.04$, corresponding to a flux above 100 MeV of  
% $6.5\times 10^{-8}$ ph cm$^{-2}$ s$^{-1}$.
% These different values testify the large variability of the source,
% especially considering that they are time averages over a rather long time.

At the beginning of 2013 the source underwent a major flare, reaching, on Jan 13, 2013, 
a flux above 100 MeV of $(110\pm 20)\times 10^{-8}$ ph cm$^{-2}$ s$^{-1}$,
more than 15 times the average photon flux listed in the 2LAC catalog (Tanaka 2013).
Most interestingly, as noted in the telegram by Tanaka (2013), the spectral
index became harder, $\Gamma=1.78\pm 0.13$. 
This implies a rising spectrum in $\nu F_\nu$, which is instead typical of
low power blazars, namely high energy peaked BL Lac objects.
Since the spectrum hardened, the luminosity variation between Jan 13 2013 and
the average value in the 2LAC catalog was almost a factor 30.
The spectral change was accompanied by a flaring behavior 
not only in $\gamma$--rays, but also in IR, optical and UV
(Carrasco et al. 2013; Donato et al. 2013),
where the source increased its flux by almost 2 orders of magnitude,
and in X--rays (Donato et al. 2013), albeit with a more modest increase.

Although other FSRQs with a relatively hard $\gamma$--ray spectrum
have been already detected by {\it Fermi}
(discussed in Padovani, Giommi \& Rau 2012 and Ghisellini et al. 2012),
it is the first time that we can well document, in a given FSRQ (i.e. a blazar with
relatively strong broad emission lines), the change from 
a soft ($\Gamma_{\rm LAT}>2$) to a hard ($\Gamma_{\rm LAT}<2$) $\gamma$--ray slope,
accompanied by an intense flare of the source from the near IR
to the $\gamma$--ray band. 
Previous hardening of the $\gamma$--ray spectrum have been observed in 
the FSRQ 4C +21.35 (=1222+216, Tanaka et al. 2011, interpreted by Foschini et al. 2011 as
a change in the location of the dissipation region in the same source) and in BL Lac
(Bloom et al. 1997).

% -------------------------------------------------------------------------
\begin{table} 
\centering
\begin{tabular}{lllll}
\hline
\hline
Catalog/Date &$F_{-8}$ [$>$0.1 GeV] &$\Gamma_{\rm LAT}$ &$\log L_\gamma$\\        
\hline   
LBAS        &$10.3\pm 1.3$ &$2.42\pm0.12$ &46.89 \\
1LAC        &$3.5$         &$2.37\pm0.10$ &46.44 \\
2LAC        &$6.5$         &$2.19\pm0.04$ &46.79 \\
Jan 13 2013 &$110\pm20$    &$1.78\pm0.13$ &48.25 \\
\hline
\hline 
\end{tabular}
\caption{
Fluxes $F_{-8}$ in units of $10^{-8}$ ph cm$^{-2}$ s$^{-1}$ above 100 MeV and corresponding spectral indices
as reported in different catalogs, and in Tanaka et al. (2013) for the Jan 13, 2013 data.
The last column reports the K--corrected [0.1--100 GeV] luminosities. 
}
\vskip -0.3 cm
\label{summary}
\end{table}
% -------------------------------------------------------------------------

The aim of this letter is to analyze the available data concerning the flare,
to compare them to the existing older observations, and to construct a coherent
picture of what caused the ``red to blue" transition of this blazar.

We use a cosmology with $h=\Omega_\Lambda=0.7$ and $\Omega_{\rm M}=0.3$.
%and use the notation $Q=Q_X 10^X$ in cgs units (except for the black hole masses,
%measured in solar mass units).

\section{Data analysis}

\subsection{{\it Swift} observations}

\emph{Swift} XRT and UVOT data were analyzed by using the {\tt HEASoft v. 6.13} software package with 
the {\tt CALDB} updated on 21 January 2013. 
XRT data were processed with {\tt xrtpipeline v. 0.12.6} with standard parameters. 
The extracted count spectra have been grouped to have at least 25 counts 
per bin, in order to adopt the $\chi^2$ test and analyzed with 
{\tt xspec v. 12.8.0} in the 0.3--10 keV energy band. 
The basic adopted model was a redshifted power law with a fixed Galactic absorption 
column of $N_{\rm H}=1.64\times 10^{20}$~cm$^{-2}$ (Kalberla et al. 2005). 
However, in the 2013 Jan 14 observation (obsID 00038401009) a broken power law model 
is required at 99.53\% according to the {\it ftest}. See Tab. \ref{swift}.

Optical/UV photometry was done with UVOT, by adopting a $5''$--sized region for the 
source and a $7''$--$60''$ source-free annulus for the background. UVOT tasks 
{\tt uvotimsum} and {\tt uvotsource} were used to perform the analysis. 
The observed magnitudes, reported in Tab. \ref{swift}, were then corrected for the 
Galactic absorption according to the filter--specific formulas of 
Cardelli et al. (1989) and converted into physical units flux by using the 
calibration of Poole et al. (2008) and Breeveld et al. (2010).

% ------------------------------------------------------------------------------------

\subsection{{\it Fermi}/LAT observations}

Publicly available {\it Fermi} LAT data were retrieved from the Fermi Science Support
Center (FSCC) and analyzed by means of the LAT Science
Tools v. 9.27.1, together with the Instrument Response
Function (IRF) Pass 7 and the corresponding isotropic and
Galactic diffuse background models. 
Source (class 2) photons in the 0.1--100
GeV energy range, collected on January 11--14 and coming from direction within 10$^\circ$
from the nominal source position were selected and filtered through standard
quality cuts. 
Standard analysis steps compliant with the FSSC recommendations
were then performed. 
Besides the target and backgrounds, all the  2FGL
point sources in the field were iwncluded in the model. 
PMNJ2345--1555 was detected with high statistical significance (TS = 340, Mattox et al. 1996) and
a high flux [$F_{0.1-100\, {\rm GeV}}=(8.76\pm 0.16) \times 10^{-7}$ photons cm$^{-2}$ s$^{-1}$] 
and a flat photon index $\Gamma=-2.02\pm0.01$. 
A significant spectral break at $E_{\rm b} = 2\pm 0.5$ GeV is found fitting the 
data with a broken power law model, with low and high spectral indices 
$\Gamma_1=1.84\pm0.15$ and $\Gamma_2=2.58\pm0.34$.
A similar analysis, but made in logarithmically spaced energy bins, gives  
results consistent with this break (solid black points in Fig. \ref{sed}). 
This suggests that the high energy hump in the SED peaks at $\sim$2 GeV.
% (corresponding to $\nu_{\rm C}$).

\section{Spectral energy distribution and model}

\begin{figure}
\vskip -0.6cm
\hskip -0.3 cm
\psfig{figure=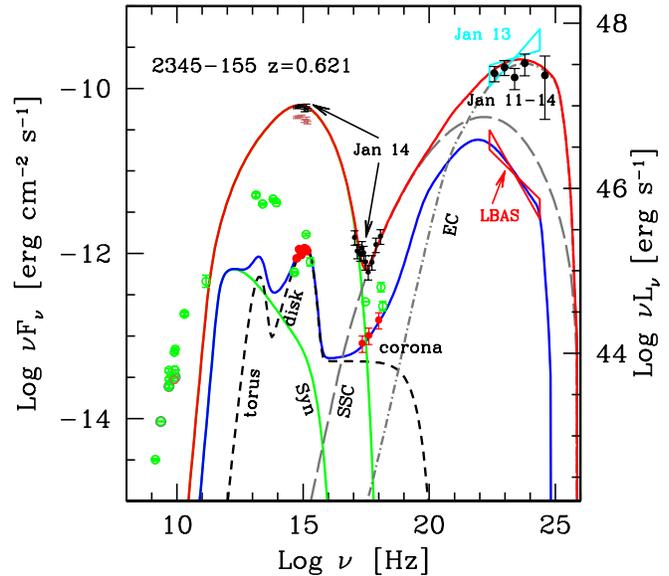,width=9cm,height=9cm}
\vskip -0.5 cm
\caption{The SED of PMN J2345--1555 and the fitting models.
Black points correspond to {\it Swift}/UVOT and XRT data
taken on Jan. 14, 2013 and to {\it Fermi} data 
integrating the flux between Jan 11 and 14.
The brown points are the UVOT data taken on Jan 17, 2013.
The cyan bow--tie corresponds to the maximum seen by
{\it Fermi} on Jan 13, 2013.
We can compare the high state of Jan 2013 with the 
low state observed in Dec 23, 2008 by {\it Swift} (red points),
while the red bow--tie refer to the average state of the first
3 months of {\it Fermi}) operations (LBAS).
Green symbols refer to archival data.
We show the models for the two states, as explained in the text and
with parameters listed in Tab. \ref{para},
with the different components labeled.
}
\label{sed}  
\end{figure}

Fig. \ref{sed} shows the SED of the source, and highlights
the dramatic flux variations occurred from the IR
to the $\gamma$--ray bands.
The cyan bow--tie corresponds to the maximum seen by {\it Fermi} on Jan 13, 2013;
black symbols correspond to quasi simultaneous data 
taken between 11 and 14 Jan ({\it Fermi}), Jan 14 (X--rays and optical--UV by {\it Swift}).
The brown symbols refer to Jan 17, 2013. 
The red symbols refer to two {\it Swift} observations merged together 
taken on Dec. 23, 2008 and Jan 10, 2009.
The red bow tie in $\gamma$--rays is the average spectrum observed in the first 3 months
of operations of {\it Fermi}.

% ----------------------------------------------------------
\begin{table}
\centering
\begin{tabular}{llll}
\hline
\hline
~                       &23/12/2008$^a$ &14/01/2013           &17/01/2013    \\
\hline
$\Gamma_{\rm single}$   &1.6$\pm$0.4    &...             &1.9$\pm$0.2       \\
$\Gamma_{\rm soft}$     &...            &2.6$\pm$0.3          &...      \\
$\Gamma_{\rm hard}$     &   &1.0$^{+0.5}_{-0.7}$  &      \\
$E_{\rm break}$ (keV)   &...            &1.7$^{+0.7}_{-0.4}$  &...    \\
$\chi^2{\rm /dof}$      &0.3/1          &3.4/6                &2.86/8     \\
$F_{\rm 0.2-10}$        &0.39$\pm$0.05  &4.6                  &2.6      \\
\hline
$V$   &18.50$\pm$0.10    &13.93$\pm$0.03 &14.23$\pm$0.03  \\
$B$   &18.61$\pm$0.06    &14.28$\pm$0.03 &14.59$\pm$0.03  \\    
$U$   &17.91$\pm$0.05    &13.40$\pm$0.03 &13.71$\pm$0.03  \\        
$W1$  &17.57$\pm$0.07    &13.39$\pm$0.04 &13.71$\pm$0.04  \\        
$M2$  &17.66$\pm$0.05    &13.32$\pm$0.04 &13.71$\pm$0.04  \\      
$W2$  &17.78$\pm$0.04    &13.49$\pm$0.04 &13.88$\pm$0.04  \\       %  &$A_V$  \\
\hline
\hline
\end{tabular}
\caption{
Top: Results of the X--ray analysis. For all spectra a fixed Galactic $N_{\rm H}=1.64\times 10^{20}$ cm$^{-2}$
was assumed. The flux $F_{\rm 0.2-10}$ is the unabsorbed flux in the [0.2--10 keV] band
in units of 10$^{-12}$  erg cm$^{-2}$ s$^{-1}$.
Bottom: UVOT Observed magnitudes.
{\it a:} Average of 2 observations on 23/12/2008 and 10/01/2009.
}
\label{swift}
\end{table}

%--------------------------------------------------
%----------------------------------------------------

One of the most interesting feature of this SED concerns the X--ray spectrum:
being a broken power law, it is strongly suggesting that in this band
we see the end of the steep synchrotron component and the emergence
of (presumably) the flat inverse Compton (IC) contribution.
This is a defining property (Padovani \& Giommi 1995) of intermediate BL Lacs
(see Tagliaferri et al. 2000 for a similar spectrum in the BL Lac source
ON 231).
Since in the previous {\it Swift}/XRT observation the entire
X--ray spectrum was flat, and could be interpreted as IC,
we must conclude that the peak of the
synchrotron flux shifted towards higher frequencies.
As a result, also the IC bump is
``bluer", making the
{\it Fermi}/LAT spectrum flat, and peaking
at $h\nu_{\rm C}\sim$2 GeV.
At the same time, the UVOT data are consistent with a
synchrotron peak at $\nu_{\rm S}\sim 10^{15}$ Hz.

If the IC process with seed photons of frequency $\nu_{\rm ext}$ produced
externally dominates (External Compton, EC for short), we have (see Tavecchio \& Ghisellini 2008):
\begin{equation}
\nu_{\rm C}\, \sim {4\over 3}\gamma_{\rm peak}^2 \nu_{\rm ext} {2 \Gamma\delta\over 1+z}
\, \to \, \gamma_{\rm peak} \sim 
\left[ {3(1+z)\nu_{\rm C} \over 8 \Gamma\delta \nu_{\rm ext} } \right]^{1/2}
\end{equation}
where $\Gamma$ is the bulk Lorentz factor; $\delta=1/[\Gamma(1-\beta\cos\theta_{\rm v}]$;
$\theta_{\rm v}$ is the viewing angle and $\gamma_{\rm peak}m_{\rm e}c^2$ is the
energy of the electrons emitting at the peaks of the SED. 
Setting $\nu_{\rm C} = 4.8\times 10^{23}$ Hz (i.e. 2 GeV) and $\Gamma=\delta$,
we derive $\gamma_{\rm peak}\sim$ $700(\Gamma/15)$  or $7000(\Gamma/15)$
according if $\nu_{\rm ext}$ is the Hydrogen Ly$\alpha$ or the peak of the IR emission 
produced by the torus, assumed to be at $3\times 10^{13}$ Hz.
The synchrotron peak frequency $\nu_{\rm S}\sim 10^{15}$ Hz then yields:
\begin{equation}
\nu_{\rm S}\, =\, 3.6\times 10^6 {\gamma_{\rm peak}^2 B \delta\over 1+z}\,\, {\rm Hz}
\, \to \, B \sim 
  { 8\Gamma \nu_{\rm ext}\nu_{\rm S} \over 1.1\times 10^7\nu_{\rm C}}  \,\,\, {\rm G}
\end{equation}
This gives 
$B\sim 70(\Gamma/15)$ or $0.7(\Gamma/15)$ G if the seed photons are coming
from the broad line region (BLR) or from the IR torus, respectively.
The very same relations hold  for the SED observed when the $\gamma$--ray 
spectrum is steep, but with different $\nu_{\rm S}$ and $\nu_{\rm C}$. 
% (as long as the synchrotron peak occurs in the thin part of the spectrum).
As we shall see, a coherent scenario explains the large shift in peak frequencies
as due to the dissipation region, usually located within the BLR, moving beyond it.
In such a case the reduced cooling allowed the electrons to reach larger $\gamma_{\rm peak}$, 
corresponding to larger $\nu_{\rm S}$ and $\nu_{\rm C}$.
%  even if the magnetic field is likely smaller at larger distances
% (i.e. $B\propto R^{-1}$).
We then apply a model along these lines.

\subsection{The model}

We use the model described in detail in Ghisellini \& Tavecchio (2009), 
and used in our previous blazar studies.
For self consistency, and to explain the main parameters listed in Tab. \ref{para},
we repeat here the main properties of the model.
 
The model accounts for several
contributions to the radiation energy density, and how these
and the magnetic one scale with the distance $R_{\rm diss}$ from the black hole 
of mass $M$.
We consider radiation from the disk 
(i.e. Dermer \& Schlickeiser 1993),
the BLR (e.g. Sikora, Begelman \& Rees 1994), a dusty torus
(see  B{\l}azejowski et al. 2000; Sikora et al. 2002), the host galaxy light
and the cosmic background radiation.
It is a one--zone, leptonic model.
% It is assumed that there is an injection of relativistic particles
% throughout the source, and that the received emission corresponds
% to a time $R/c$ since the start of the injection.
% Even if injection lasted longer,
% adiabatic losses caused by the expansion of the source (which
% is traveling while emitting) and the corresponding decrease 
% of the magnetic field would make the observed flux to decrease. 
% Therefore our calculated spectra correspond to the maximum of
% a flaring episode.

The emitting region is spherical, of size $R\sim0.1 R_{\rm diss}$.
% moving with a bulk Lorentz factor $\Gamma$ 
% and located at a distance 
% $R_{\rm diss}$ from the black hole of mass $M$.
% The bolometric luminosity of the accretion disk is $L_{\rm d}$.
The jet accelerates in its inner parts with $\Gamma\propto R_{\rm diss}^{1/2}$ 
% ($d$ is the distance from the black hole), 
up to a value $\Gamma_{\rm max}$. 
% In the acceleration region the jet is parabolic 
% (following, e.g. Vlahakis \& K\"onigl 2004) and beyond this point the jet becomes
% conical with a semi--aperture angle $\psi$ (assumed to be 0.1).

% The energy particle distribution $N(\gamma)$ [cm$^{-3}$]
% is calculated solving the continuity 
% equation where particle injection, radiative cooling and pair production
% (via the $\gamma$--$\gamma \to e^\pm$ process) are taken into account.
The electron injection function $Q(\gamma)$ [cm$^{-3}$ s$^{-1}$]
is assumed to be a smoothly joining broken power--law,
with a slope $Q(\gamma)\propto \gamma^{-{s_1}}$ and
$\gamma^{-{s_2}}$ below and above a break energy $\gamma_{\rm b}$:
%
% \begin{equation}
% Q(\gamma)  \, = \, Q_0\, { (\gamma/\gamma_{\rm b})^{-s_1} \over 1+
% (\gamma/\gamma_{\rm b})^{-s_1+s_2} }
% \label{qgamma}
% \end{equation}
%
The total power injected into the source in relativistic
electrons is $P^\prime_{\rm i}=m_{\rm e}c^2 V\int Q(\gamma)\gamma d\gamma$,
where $V=(4\pi/3)R^3$. 
% is the volume of the emitting region.

\begin{figure}
%\vskip -0.4cm
\hskip -0.3cm
\psfig{figure=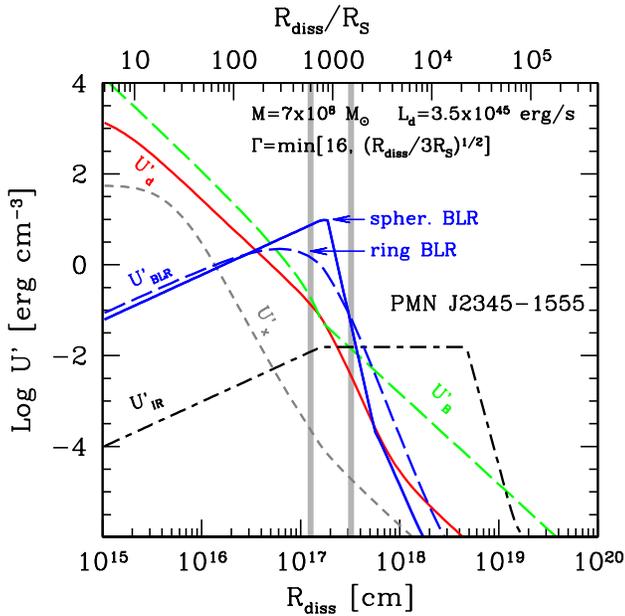,width=9cm,height=9cm}
\vskip -0.5  cm
\caption{The energy density measured in the comoving frame of the blob as a function
of the distance form the black hole, $R_{\rm diss}$.
We show separately the contribution of the magnetic field ($U_B$), 
of the radiation coming directly from the disk ($U^\prime_{\rm d}$),
and its X--ray corona ($U^\prime_X$), of the radiation produced by the BLR  ($U^\prime_{\rm BLR}$)
and by the torus ($U^\prime_{\rm IR}$).
The two vertical grey lines mark the values $R_{\rm diss}$ for the 
two states of the source. 
The Jan 13, 2013 flare corresponds to dissipation occurring just beyond the BLR.
For illustration, we also show $U^\prime_{\rm BLR}$ for a ``ring--like" BLR lying in the disk,
shaped as a torus of radius $R_{\rm BLR}$ and cross sectional radius $R_{\rm BLR}/10$.
%This and the spherical geometry should bracket the other possibilities.
}
\label{u2345}  
\end{figure}

The BLR, reprocessing 10\% of the disk luminosity $L_{\rm d}$, is assumed to be a thin spherical shell 
located at a distance $R_{\rm BLR}=10^{17}L_{\rm d, 45}^{1/2}$ cm.
%Since $R_{\rm BLR}\propto L_{\rm d}^{1/2}$, the
The radiation energy density of the broad lines is constant within the BLR, but 
it is seen amplified by $\sim\Gamma^2$ by the moving blob  (as long
as $R_{\rm diss}<R_{\rm BLR}$).
For illustration, Fig. \ref{u2345} shows also the case of a ``ring--like"
BLR, lying in the disk, shaped as a torus of radius $R_{\rm BLR}$ and
cross sectional radius equal to $0.1 R_{\rm BLR}$. 
The $U^\prime_{\rm BLR}$ profile is somewhat smoother in this case. 
The two cases (spherical and ring--like) 
should bracket the possible geometrical possibilities.
A dusty torus, located at a distance 
$R_{\rm IR}=2.5\times 10^{18}L_{\rm d, 45}^{1/2}$ cm,
reprocesses $\sim$30\% of $L_{\rm d}$
%through dust emission 
in the far IR.
In the inner parts of the accretion disk there is an X--ray emitting
corona of luminosity $L_{\rm X}\sim 0.3L_{\rm d}$
with a spectrum  $F(\nu) \propto \nu^{-1}\exp(-h\nu/150\,{\rm keV})$.

The energy densities of 
all these external components are calculated in the jet comoving frame, and
used to calculate the resulting EC spectrum. 
The internally produced synchrotron emission is used to calculate the synchrotron
self Compton (SSC) flux.

We adopt a standard Shakura \& Sunjaev (1973) accretion disk spectrum.
It depends on the black hole mass $M$ and the accretion rate $\dot M$, that can be
found if the SED shows signs of disk emission.
In this case the total disk luminosity $L_{\rm d}$ fixes $\dot M$, and the
peak frequency of the disk spectrum fixes $M$.
Furthermore, we are helped by the presence of broad lines, that can be used as a proxy for
$L_{\rm d}$. 
This method, detailed in Calderone et al. (2013), returns very accurate
black hole masses  
% (better than a factor 2)
if the peak of the disk emission is visible.
% In this cases it is therefore more accurate than the widely used virial
% methods, based on the FWHM of the broad lines and to scaling relation
% between the ionizing luminosity and the radius of the BLR.

%--------------------------------------------------
\begin{figure}
\vskip -0.5cm
\hskip -0.3 cm
\psfig{figure=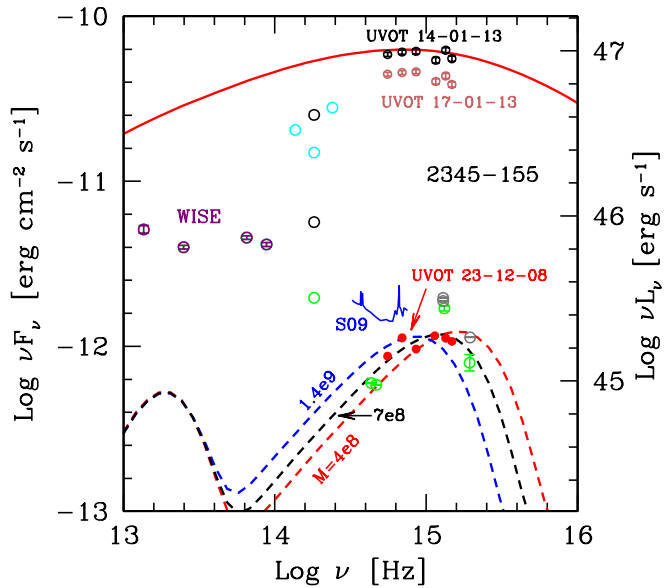,width=9cm,height=9cm}
\vskip -0.5 cm
\caption{
Zoom of the SED of PMN J2345--1555 to better show the
IR--UV part of the spectrum. 
The {\it Swift}/UVOT data of Dec. 23, 2008 can be described
by a standard accretion disk model.
Note the upturn in the optical spectrum taken by Shaw et al. (2009),
labeled S09.
To illustrate the possible uncertainties on the black hole mass,
we show three accretion disk (+torus) models that have the 
same disk luminosity but different black hole masses,
as labelled. 
In Jan 13, 2013 the optical--UV synchrotron flux increased by almost 2
orders of magnitude, largely overtaking the thermal emission.
}
\label{zoom}  
\end{figure}
%----------------------------------------------------

% --------------------------------------------------------------------------------
\begin{table*} 
\centering
\begin{tabular}{llllllllllllllll}
\hline
\hline
Date &$R_{\rm diss}$ &$R_{\rm diss}/R_{\rm S}$ &$M$ &$R_{\rm BLR}$ &$P^\prime_{\rm i}$ &$L_{\rm d}$ &$L_{\rm d}/L_{\rm Edd}$ &$B$ &$\Gamma$ &$\theta_{\rm v}$
     &$\gamma_{\rm b}$ &$\gamma_{\rm max}$ &$\nu_{\rm S}$ &$s_1$  &$s_2$ \\ % &$\gamma_{\rm c}$ &$\gamma_{\rm peak}$ &$U\prime$ \\
~[1]      &[2] &[3] &[4] &[5] &[6] &[7] &[8] &[9] &[10] &[11] &[12] &[13] &[14]  &[15]  &[16]\\
\hline   
23/12/2008$^a$ &132 &1100 &4e8  &190  &3.5e--3 &3.6 &0.06  &1.8  &13 &3    &100   &4e3 &1.e12  &--1 &2.8 \\ % &11   &92.71 &5.0 \\
23/12/2008     &126 &600  &7e8  &186  &1.6e--3 &3.5 &0.033 &1.9  &13 &2.4  &180   &5e3 &2.7e12 &0.2 &2.9 \\ % &11.5 &141   &5.0  \\
13/01/2013     &336 &1600 &7e8  &186  &0.014   &3.5 &0.033 &0.57 &16 &2.4  &7e3   &5e4 &1.4e15 &0   &3.2  \\% &292  &5485  &0.073 \\
\hline
\hline 
\end{tabular}
% \vskip 0.4 true cm
\caption{
Parameters of the models shown in Fig. \ref{sed} and Fig. \ref{zoom}.
$a$: set of parameters used in Ghisellini et al. (2010), where a slightly smaller
$M$ was adopted.
Col. [1]: date; 
Col. [2], [3]: dissipation radius in units of $10^{15}$ cm and in units of $R_{\rm S}$;
Col. [4]: black hole mass in solar masses;
Col. [5]: size of the BLR in units of $10^{15}$ cm;
Col. [6]: power injected in the blob calculated in the comoving frame, in units of $10^{45}$ erg s$^{-1}$; 
Col. [7], [8]: accretion disk luminosity in units of $10^{45}$ erg s$^{-1}$ and
         in units of $L_{\rm Edd}$;
Col. [9]: magnetic field in Gauss;
Col. [10]: bulk Lorentz factor at $R_{\rm diss}$;
Col. [11]: viewing angle in degrees;
Col. [12], [13]: break and maximum random Lorentz factors of the injected electrons;
Col. [14]: peak synchrotron frequency $\nu_{\rm S}$ (Hz, rest frame);
Col. [15] and [16]: slopes of the injected electron distribution [$Q(\gamma)$]. 
% below and above $\gamma_{\rm b}$.
}
\label{para}
\end{table*}
% --------------------------------------------------------------------------------

%--------------------------------------------------
%----------------------------------------------------

\section{Results}

Fig. \ref{sed} shows the result of the modeling:
the red solid curve is the model for the data of Jan 13, 2013, and
the solid blue line is the model fitting the Dec. 2008 UVOT and
XRT data, together with the average $\gamma$--ray flux of the
first 3 months of {\it Fermi}.
Both models share the same black hole mass, accretion rate and viewing angle.
All used parameters are listed in Tab. \ref{para} that also reports, for
ease of the reader, the set of parameters used in Ghisellini et al. (2009),
that considered a slightly smaller black hole mass.
Fig. \ref{sed} shows also the torus, disk and corona emission
(unchanged for both states), the synchrotron component
and the SSC and EC contributions for the Jan 13 SED.
% The two IR--$\gamma$--ray SEDs (Jan 13 2013 and Dec. 2008) differs by a factor close to 100.
Particularly important is the UVOT spectrum in Dec 2008, that can be interpreted
as the peak of the accretion disk component.
The Shaw et al. (2009) optical spectrum is visible in Fig. \ref{zoom}.
Its flux is larger than in Dec. 2009, there is a contamination from the synchrotron
component at low frequencies, an upturn at large frequencies and the broad MgII and H$\beta$
lines are well visible. 
Their luminosities can be used to reconstruct the luminosity of the entire BLR
(using the template of Francis et al. 1991 or Vander Berk et al. 2001)
and then $L_{\rm d}$.
It agrees with the ones we have adopted, i.e. $L_{\rm d}\sim 3.5\times 10^{45}$ erg s$^{-1}$.
Since the UVOT points show a peak in $\nu F_\nu$, we find $M=7\times 10^8 M_\odot$.
To show how well the black hole mass can be determined, we have
plotted in Fig. \ref{zoom} other two disk spectra, of the same $L_{\rm d}$ but
with different masses (i.e. 0.4 and 1.4 billion solar masses).
With these values the data are not well reproduced.
By applying the virial method, Shaw et al. (2009) estimated $M\sim(3-6)\times 10^8M_\odot$,
using the H$\beta$ and the MgII lines.
Given the uncertainties, these values are well consistent with ours.

To reproduce the two states of the source, we have chosen to modify
a minimum numbers of parameters.
Besides $M$ and $\dot M$, the models share the same magnetic field
profile (i.e. the same Poynting flux $L_B\equiv R^2 \Gamma^2 B^2c/8$;
see Tab. \ref{powers}).
The main changes concern the typical energies of the injected electrons,
higher in the high states, the value of $R_{\rm diss}$ and the injected power.
We have furthermore assumed that in the high state the bulk
Lorentz factor is somewhat larger ($\Gamma$=16 vs. 13).
As shown in Fig. \ref{u2345}, all these changes are consistent with the
idea that the two states correspond to 
a different $R_{\rm diss}$: for the ``low" state, 
$R_{\rm diss}<R_{\rm BLR}$, the radiative cooling is strong,
electrons cannot reach high energies and the overall SED is ``red".
By moving from 600 to 1600 \sc\ radii, the radiation energy density drops,
letting the electrons to reach higher energies.
The magnetic field also decreases [$B\propto (R_{\rm diss}\Gamma)^{-1}$], but
the change of the radiation energy density is more drastic,
making the Compton dominance (ratio of the inverse Compton to synchrotron
luminosities) to decrease in the high state.

%----------------------------------------------------
\begin{table} 
\centering
\begin{tabular}{lllll}
\hline
\hline
Date &$\log P_{\rm r}$ &$\log P_{\rm B}$ &$\log P_{\rm e}$ &$\log P_{\rm p}$ \\
\hline   
23/12/2008$^a$  &44.72 &44.56 &43.97 &45.92  \\  
23/12/2008      &44.39 &44.55 &43.61 &45.62   \\    
13/01/2013      &45.53 &44.54 &44.67 &45.25 \\
\hline
\hline 
\end{tabular}
% \vskip 0.4 true cm
\caption{Jet power in the form of radiation, Poynting flux,
bulk motion of emitting (therefore relativistic) electrons and cold protons, 
assuming one proton per emitting electron.
$a$: set of powers derived in Ghisellini et al. (2010), where
a slightly smaller black hole mass was adopted.
}
\label{powers}
\end{table}
%--------------------------------------------------

\section{Discussion}

\begin{figure}
\vskip -0.6cm
\hskip -0.7cm
\psfig{figure=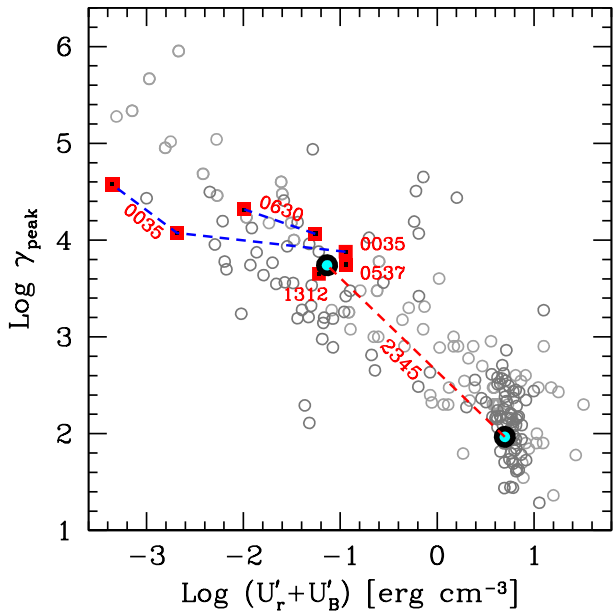,width=9.5cm,height=9.5cm}
\vskip -0.5  cm
\caption{
Random Lorentz factor ($\gamma_{\rm peak}$) of the electrons emitting at
the SED peaks as a function
of the total energy density $U^\prime=U'_{\rm r}+U^\prime_B$
as measured in the comoving frame.
Small $\gamma_{\rm peak}$ and large $U^\prime$ 
correspond to ``red" blazars, while large 
$\gamma_{\rm peak}$ and small $U^\prime$ correspond to ``blue" blazars.
The grey points correspond to blazar analyzed by us in the past
(Ghisellini et al., 2012 and references therein),
while red symbols refer to the 4 ``blue" quasars studied in
Padovani, Giommi \& Rau (2012) and Ghisellini et al. (2012).
PMN J2345--1555 is the first clear example of a blazar
shifting from red to blue.
}
\label{gpeak}  
\end{figure}

Fig. \ref{gpeak} shows $\gamma_{\rm peak}$ as a function of the total radiation
energy density (magnetic plus radiative) as seen in the comoving frame of the source.
The random Lorentz factor $\gamma_{\rm peak}$ corresponds to the energy of the electrons
emitting most of power (both in synchrotron and in inverse Compton).
The grey points are the values of blazars analyzed by us in the past (see
Ghisellini et al. 2012 and references therein), while the 
red squares corresponds to the 4 ``blue" blazars discovered by Padovani, Giommi \& Rau (2012).
As these authors suggested, they are very likely FSRQs with a 
a very powerful optical synchrotron emission, that hides the emission lines.
They should therefore correspond to the high state of PMN J2345--1555.
The monitoring of these 4 blazars is too sparse to know if they are permanently
``in the high state", or if we caught them by chance.
In any case, PMN J2345--1555 demonstrates that a single source can indeed 
vary not only its overall flux by orders of magnitudes, but also 
its overall ``look" and ``color".
In our scheme this this is simply due to the change of the location of the
dissipation region by a factor less than 3 (from 600 to 1600 \sc\ radii in this case).
This is sufficient to shift a blazar, in Fig. \ref{gpeak}, from the red
($\gamma_{\rm peak}\lsim 10^3$ and $U^\prime\gsim 0.3$ ) to the
blue zone ($\gamma_{\rm peak}\gsim 10^3$ and $U^\prime\lsim 0.3$ erg cm$^{-3}$).
The top left region of the plane is populated mainly by low power, high energy peaked
BL Lacs, lacking strong broad emission lines. 
For them, we do not expect strong variations of the cooling process 
when $R_{\rm diss}$ changes (since there is no BLR). 
Variations are however possible due to changes in the acceleration process,
in turn linked to the total power.
Larger injected powers (and thus larger observed luminosities)
could be associated to larger $\gamma_{\rm peak}$,
as observed in Mkn 501 (Pian et al. 1998; Tavecchio et al. 2001).

%--------------------------------------------------
%----------------------------------------------------

The fact that PMN J2345--1555 is the first FSRQ "caught in the act"
suggests that it is a rare phenomenon.
Furthermore, out of a sample of hundreds of FSRQs
detected by {\it Fermi}, only 4 were found to be blue
by Padovani et al. (2012).
This leads us to conclude that only a few per cent of
FSRQs dissipate most of their energy beyond the BLR,
either occasionally or permanently.
In turn, this suggests that the region where the 
jet dissipates the most is indeed well confined at
$R_{\rm diss} \lsim 10^3 R_{\rm S}$.
Larger radii (even by a small amount), corresponding
to a blue FSRQ, are rare, although not impossible.

We envisage two possibilities to have a dissipation always well localized.
The first is internal shocks: two shells initially separated by $\Delta R$
and moving with different $\Gamma$ will collide at
$R_{\rm diss}\sim \Gamma^2 \Delta R$.
If $\Delta R =$ a few $R_{\rm S}$, and $\Gamma\sim 15$, then
$R_{\rm diss}\sim 10^3 R_{\rm S}$.
Larger values are possible for larger $\Gamma$ or for larger $\Delta R$.
Alternatively, $R_{\rm diss}$ may be set by the pressure balance 
between the jet and the external medium.
In this case  $R_{\rm diss}$ is larger for a larger jet pressure, and then 
for a larger jet power.

\section*{Acknowledgments}
FT and GB acknowledge financial contribution from a PRIN--INAF--2011 grant.
We acknowledge the use of public data from the {\it Swift} data archive. 
This research made use of the NASA/IPAC Extragalactic Database (NED) 
which is operated by the Jet Propulsion Laboratory, Caltech,
under contract with the NASA, and of data obtained from the High Energy Science Archive 
Research Center (HEASARC), provided by NASA's GSFC, and 
of the XRT Data Analysis Software (XRTDAS) developed under the responsibility of the 
ASI Science Data Center (ASDC), Italy.


\begin{thebibliography}{}

\bibitem[]{} Abdo A.A.
% , Ackermann M., Ajello M., 
et al., 2009, ApJ, 700, 597 % LBAS 3 months

\bibitem[]{} Abdo A.A.
% , Ackermann M., Ajello M., 
et al., 2010, ApJ, 715, 429 % 1LAC 

\bibitem[]{} B{\l}azejowski M., Sikora M., Moderski R. \& Madejski G.M., 2000, ApJ, 545, 107

\bibitem[]{} Bloom S.D.
% , Bertsch D.L., Hartman R.C. 
et al., 1997, ApJ, 490, L145 % flare of BL Lac in EGRET

\bibitem[]{} Breeveld A.A.
% , Curran P.A., Hoversten E.A. 
et al., 2010, MNRAS, 406, 1687  

\bibitem[]{} Calderone G., Ghisellini G., Colpi M. \& Dotti M., 2013, MNRAS, in press
             (astro--ph/1212.1181)

\bibitem[]{} Cardelli J.A., Clayton G.C. \& Mathis J.S., 1989, ApJ, 345, 245

\bibitem[]{} Carrasco L., Recillas E., Mayya D.Y. \& Carraminana A., 2013, 
             ATel 4736 %  H = 12.231 +/- 0.11 Dec 18, 2012
             
% \bibitem[]{} Carrasco L., Recillas E., Escobedo G. \& Carraminana A., 2013, ATel 4608
        % H=12.725+/-0.06 on Nov 26, 2012

% \bibitem[]{} Celotti A. \& Ghisellini G., 2008, MNRAS, 385, 283  
             
\bibitem[]{} Dermer C.D. \& Schlickeiser R., 1993, ApJ, 416, 458

\bibitem[]{} Donato D., Cheung C.C. \& Tanaka Y., 2013, ATel 4742 % swift XRT & Uvot

\bibitem[]{} Foschini L., Ghisellini G., Tavecchio F., Bonnoli G. \& Stamerra A., 2011,  
             Fermi Symposium proceedings eConf C110509 (astro--ph/1110.4471) 

\bibitem[]{} Francis P.J., Hewett P.C., Foltz C.B., Chaffee F.H., Weymann R.J., Morris S.L., 
             1991, ApJ, 373, 465

% \bibitem[]{} Ghisellini G. \& Tavecchio F., 2008, MNRAS, 387, 1669  % a new perspective+blue

\bibitem[]{} Ghisellini G. \& Tavecchio F., 2009, MNRAS, 397, 985   % canonical

\bibitem[]{} Ghisellini G., Maraschi L. \& Tavecchio F., 2009, MNRAS, 396, L105 % divide

\bibitem[]{} Ghisellini G., Tavecchio F., Foschini L., Ghirlanda G., Maraschi L. \& Celotti A., 
             2010, MNRAS, 402, 497   % General physical properties of bright Fermi blazars
	
\bibitem[]{} Ghisellini G., Tavecchio F., Foschini L., Sbarrato T., Ghirlanda G. \& Maraschi L., 
             2012, MNRAS, 425, 1371 % blue FSRQ del Giommi
             
\bibitem[]{} Healey S.E.
% , Romani R.W., Cotter G. 
et al., 2008, ApJS, 175, 97   % z, CGRaBS

\bibitem[]{} Kalberla P.M.W., Burton W.B., Hartmann D., Arnal E.M., Bajaja E., 
             Morras R. \& P\"oppel W.G. L. 2005, A\&A, 440, 775 % AV recente

\bibitem[]{} Mattox J.R.
%, Bertsch D.L., Chiang J. 
et al., 1996, ApJ, 461, 396 

\bibitem[]{} Nolan P.L.
%, Abdo A.A., Ackermann M. 
et al. 2012, ApJS, 199, 31 % 2LAC

\bibitem[]{} Padovani P., Giommi P. \& Rau A., 2012, MNRAS, 422, L48 % blue FSRQs

\bibitem[]{} Padovani P. \& Giommi P., 1995, ApJ, 444, 567

\bibitem[]{} Pian E.
%, Vacanti G., Tagliaferri G. 
et al., 1998, ApJ, 492, L17 % Mkn 501

\bibitem[]{} Poole T.S.
%, Breeveld A.A., Page M.J. 
et al., 2008, MNRAS, 383, 627

%\bibitem[]{} Roming P.W.A.
%, Kennedy T.E., Mason K.O. et al., 2005, Space Sci. Rev., 120, 95

% \bibitem[]{} Sbarrato T., Ghisellini G., Maraschi L. \& Colpi M., 2012, MNRAS, 421, 1764 % Lblr vs Lgamma

\bibitem[]{} Shakura N.I. \& Syunyaev R.A., 1973, A\&A, 24, 337

\bibitem[]{} Shaw M.S., Romani R.W., Healey S.E., Cotter G., Michelson P.F. \& Readhead A.C.S., 
             2009, ApJ, 704, 477 % optical spectrum 

\bibitem[]{} Sikora M., Begelman M.C. \& Rees M.J., 1994, ApJ, 421, 153 
      
\bibitem[]{} Sikora M., Blazejowski M.; Moderski R.,  Madejski G.M., 2002, ApJ. 577, 78
             
\bibitem[]{} Tagliaferri G.
%, Ghisellini G., Giommi P. 
et al., 2000, A\&A, 354, 431 % ON 231
             
\bibitem[]{} Tanaka Y., 2013, ATel 4735 % Fermi Jan 13 2013, F=(1.1+/-0.2)e-6, Gamma=1.78+/-0.13 
             % with respect to 2.17+/-0.04       in the 2FGL catalog
             
\bibitem[]{} Tanaka Y.T.
% , Stawarz \L., Thompson D.J. 
et al., 2011, ApJ, 733, 19 % 1222+216
        
\bibitem[]{} Tavecchio F., \& Ghisellini G., 2008, MNRAS, 386, 945 % BLR
   
\bibitem[]{} Tavecchio F.
% , Maraschi L., Pian E. 
et al., 2001, ApJ, 554, 725 % Mkn 501
         
\bibitem[]{} Vanden Berk D.E.
% , Wilhite B.C., Kron R.G. 
et al., 2001, AJ, 122, 549

\end{thebibliography}
\end{document}